\documentclass[aps,prb,twocolumn]{revtex4-1}
\usepackage[utf8]{inputenc}
\usepackage{amsmath}
\usepackage{amssymb}
\usepackage{graphicx}
\usepackage{color}
\usepackage[bookmarks=false]{hyperref}
\usepackage[usenames,dvipsnames]{xcolor}
\usepackage{subfigure}
\usepackage[pdf]{pstricks}

\def\be{\begin{equation}}       \def\ee{\end{equation}}
\def\bea{\begin{eqnarray}}      \def\eea{\end{eqnarray}}
\def\ba{\begin{array} }
\def\ea{\end{array} }
\def\bnum{\begin{enumerate} }
\def\enum{\end{enumerate}}

\def\=>{\Rightarrow}
\def\>{\rightarrow}

\def\eye2{Fathbb{I}}

\def\d0{\Delta_{0}}

\def\Co122{ BaFe$_{2-x}$Co$_x$As$_2$}

\begin{document}

\title{Pseudogap crossover  in the electron-phonon system}
\author{I.~Esterlis$^1$, S.~A.~Kivelson$^{1,2}$, D.~J.~Scalapino$^3$}
\affiliation{$^1$Department of Physics, Stanford University, Stanford, California 94305, USA \\
$^2$Geballe Laboratory for Advanced Materials, Stanford University, Stanford, CA 94305, USA \\
$^3$Department of Physics, University of California, Santa Barbara, CA 93106-9530, USA
}
\date{\today}

\begin{abstract}
Thermodynamic properties of the square-lattice Holstein model of the electron-phonon problem with phonon frequencies small compared to the bare Fermi energy are obtained using Monte Carlo methods, a strong-coupling (bipolaronic) expansion, and a weak coupling Migdal-Eliashberg approach.  Already at elevated temperatures where the  charge-density wave (CDW) and superconducting (SC) correlations are very short-range, a crossover occurs as a function of increasing  electron-phonon coupling, $\lambda_0$, from a normal metallic  regime to a pseudogap  regime. At sufficiently low $T$, a  SC phase is found for small $\lambda_0$ and a commensurate insulating CDW phase for large $\lambda_0$. 
\end{abstract}
\maketitle

\textit{Introduction} -- Electron-phonon interactions determine many of the electronic properties of quantum materials;  this includes the normal state electrical transport properties of most metals at all but the lowest temperatures, and of course the nature of the superconducting (SC) and/or charge-density-wave (CDW) ground-states of many ``conventional'' materials.  In all but a few cases, the dimensionless electron-phonon coupling constant, $\lambda_0$, is of order one.  Nonetheless, there is a genuine small parameter in the problem, the ratio of the phonon energy, $\hbar \omega_0$, to the bare Fermi energy, $E_F^{(0)}$.  While it has been argued that Migdal-Eliashberg (ME) theory\cite{migdal1958,eliashberg1960} provides an accurate solution to this problem provided $\lambda_0 \ll E_F^{(0)}/\hbar\omega_0$, we recently showed\cite{PhysRevB.97.140501, esterlis2018bound} that ME theory breaks down when $\lambda_0 \sim 1$, even when the nominal condition for its validity is satisfied. As was already  suggested in various earlier studies,
\footnote{
The breakdown of ME theory has been discussed  previously from the perspective of dynamical mean field theory\cite{PhysRevB.48.6302, PhysRevB.54.5389, PhysRevB.58.14320, PhysRevLett.89.196401, PhysRevLett.91.186405} and through an analysis\cite{0295-5075-56-1-092} of ground-state properties in the $M\to \infty$ limit.
}
this breakdown is associated with the non-perturbative  formation of bipolarons.

\begin{figure}[t!]
    \centering
     \includegraphics[]{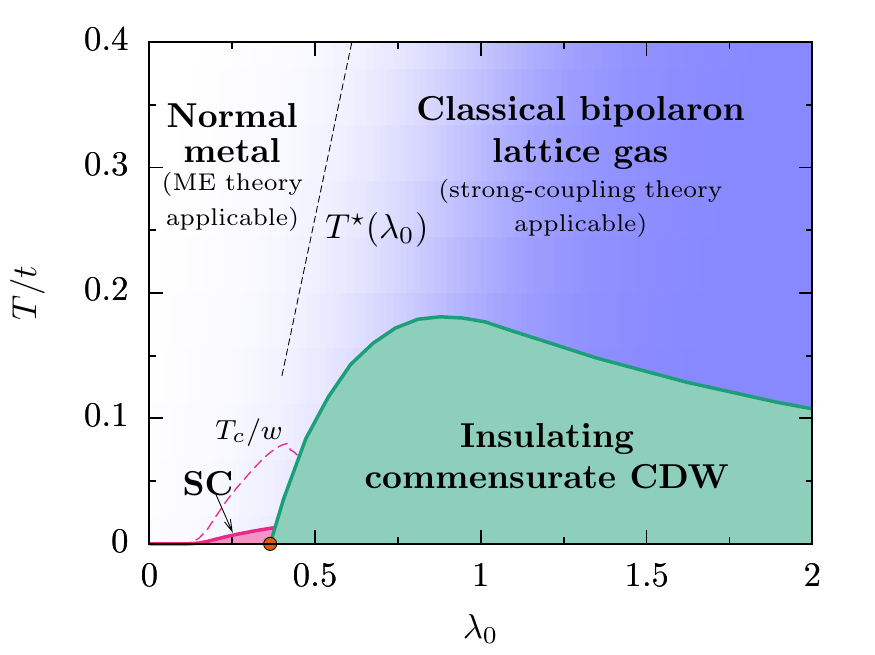}
    \caption{Schematic phase diagram of the electron-phonon problem with small $w\equiv \hbar\omega_0/E_F^{(0)}$.   Depending on details of the band-structure and the electron density, additional CDW phases can arise (including metallic ones at  intermediate $\lambda_0$).}
    \label{schematic}
\end{figure}

In the present paper, we explore the global phase diagram of the Holstein model \cite{HOLSTEIN1959325} -- the paradigmatic model of the electron-phonon problem -- over a broad range of temperatures, $T$, and $\lambda_0$ in the physically important limit $\hbar\omega_0/E_F^{(0)} \ll 1$.  We have carried out extensive Monte Carlo (MC) calculations, which we then compare with the results of ME theory and with a strong-coupling expansion (in powers of $1/\lambda_0$).  As shown in the schematic phase diagram in Fig \ref{schematic}, there are two regions separated by a crossover line, $T = T^\star(\lambda_0)$; ME theory gives a good account of the physics only in the left region while a strong-coupling ``polaronic'' approach is accurate to the right.  (Naturally, neither approach is entirely reliable close to the crossover line.)  

The physics in the two regions is correspondingly distinct:  In the weak coupling regime, the properties of the normal state are  dominated by weakly scattered quasi-particle excitations near a well-defined Fermi surface with decay rates $\hbar\gamma \sim \lambda_0 T$ and there is a low-$T$  superconducting ground-state with a transition temperature  $T_c$ which is proportional to $ \hbar\omega_0$ times a (possibly non-monotonic)  function of $\lambda_0$.  In the strong-coupling limit, there is a ``pseudogap" to single-particle excitations, the normal state is a classical lattice gas of (effectively non-dynamical) bipolarons with binding energy $\sim \lambda_0 E^{(0)}_F$ and at low temperatures the system has a tendency to commensurate CDW states, with ordering vectors unrelated to any Fermi-surface nesting vector. Depending on the electron density  there may be a sequence of transitions to higher-order commensurate states or phase separation.

We will study the Holstein Hamiltonian \cite{HOLSTEIN1959325}
\begin{align}
	H &= H_e + H_p + H_{e-p}, \label{eq:hamiltonian} \\ 
	H_e &= -\sum_{ij\sigma}t_{ij}c^\dag_{i\sigma}c_{j\sigma} -\mu\sum_{i\sigma}   n_{i\sigma}, \label{eq:he} \\
	H_p &= \sum_i \frac{p_i^2}{2M} + \frac 12 K x_i^2, \label{eq:hp} \\
	H_{e-p} &= \alpha\sum_{i\sigma} x_i n_{i\sigma}, \label{eq:hep}
\end{align}
and  $\omega_0 = \sqrt{K/M}$. The important dimensionless parameters are the coupling strength $\lambda_0 \equiv \alpha^2 \rho^{(0)}/K$ and the retardation parameter $w\equiv \hbar \omega_0/E_F^{(0)}$, where $\rho^{(0)}$ is the bare ($\alpha=0$) density of states at the Fermi energy. 
\footnote{Note that beyond the weak coupling regime, $\lambda_0 \ll 1$, there is a non-trivial relation\cite{PhysRevB.97.140501} between the bare quantities and the ``renormalized'' or ``physical'' values of $\lambda$, $\rho(E_F)$, and the phonon dispersion, $\omega_{\bf q}$.}
While there are notable exceptions, the regime of the Holstein model relevant to most materials is $ w \ll 1$.  

We therefore study the phase diagram and thermodynamic correlation functions of the Holstein model in the limit $w \ll 1$.  In previous work,\cite{PhysRevB.97.140501} we carried out such a study via determinant quantum Monte Carlo (DQMC) for $w=0.1$, but only for weak to moderate $\lambda_0$:  $0 \leq \lambda_0 \leq 0.6$. The DQMC method is challenging to employ in the strong-coupling regime due to prohibitively long autocorrelation times for temperatures much lower than $E_F^{(0)}$.\cite{hohenadler2008autocorrelations} Fortunately, in this strong-coupling regime especially, the results are not expected to depend strongly on $M$ so long as $w \ll 1$, an expectation that we have confirmed where it can be tested.
\footnote{ In Ref. \onlinecite{PhysRevB.98.085405}, a similar comparison was made between results with $w=0$ and non-zero $w$ for the half-filled Holstein model.}
Thus, to analyze the full phase diagram, we will consider here the limit $M\rightarrow \infty$, corresponding to $w \rightarrow 0$. In this limit the phonons become classical variables, and the MC calculations become substantially simpler. (See the Appendix for details of the MC algorithm employed.) 
We have chosen parameters to avoid any non-generic band features or special commensurate densities -- specifically we take the matrix $t_{ij}$ to contain both nearest-neighbor hopping $t$ and next-nearest-neighbor hopping $t'$, with the ratio $t'/t=-0.3$.  We work at a fixed chemical potential, chosen such that the density is $n=0.8$ at $T=0.25t$. We have studied systems of linear size $L \leq 12$ with periodic boundary conditions and temperatures $T \geq t/40$.

\begin{figure}[t!]
    \centering
     \includegraphics[]{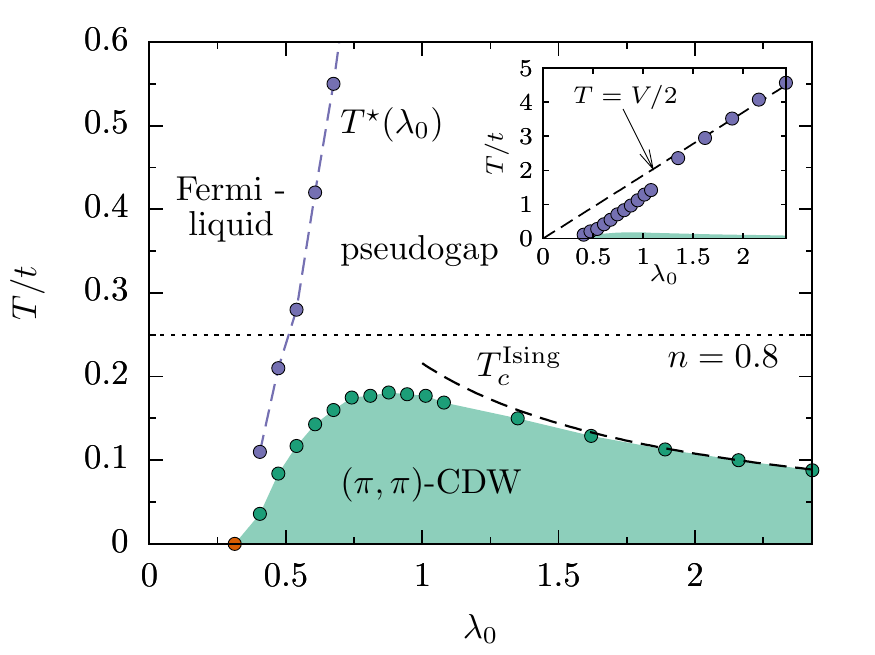}
    \caption{Phase diagram of the Holstein model in the $M\to \infty$ limit with $t'/t=-0.3$. Chemical potential has been chosen such that $n=0.8$ for $T=0.25t$. The inset shows the phase diagram extended to higher temperatures where, for $\lambda_ 0 \gtrsim 1$, $T^\star$ becomes equal to half the bipolaron binding energy, $V/2$, as expected from the strong-coupling limit. See text for additional details.}
    \label{fig:phase_diag}
\end{figure}

\begin{figure}[t!]
    \centering
    \includegraphics[]{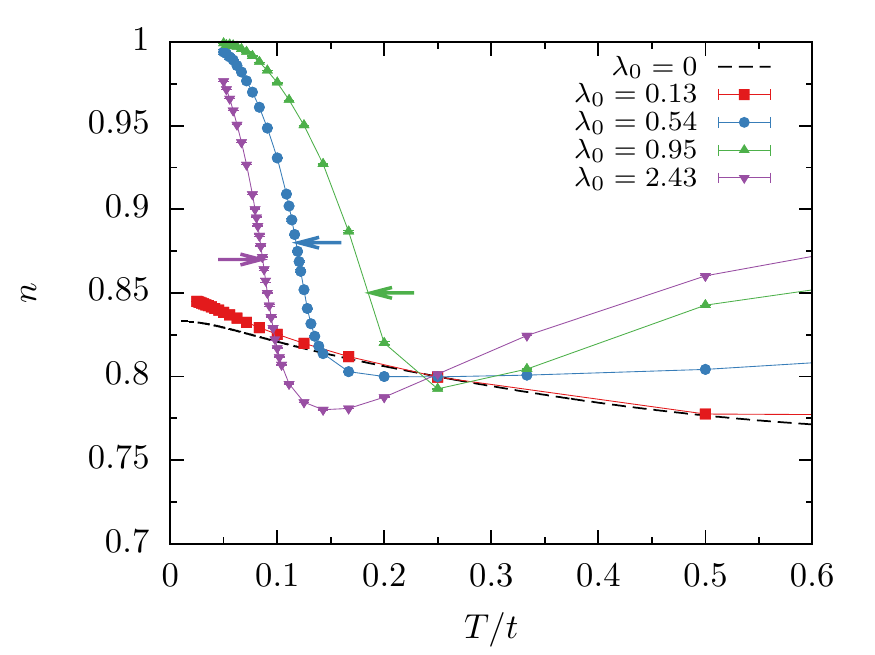}
    \caption{Density $n$ as a function of $T$ for various coupling strengths. The dashed line shows the evolution of the density for non-interacting electrons. Arrows indicate $T_\mathrm{cdw}$ inferred from finite-size scaling. Linear system size is $L=12$.}
    \label{fig:density_vs_temp}
\end{figure}

\textit{Results} -- The phase diagram derived from our MC studies in the $M\to \infty$ ($w \to 0$) limit is shown in Fig. \ref{fig:phase_diag}. For $\lambda_0 \ll 1$ we find a translationally invariant Fermi liquid ground-state. At strong-coupling $\lambda_0 \gg 1$ the low-energy degrees of freedom are bipolarons, which have a binding energy $V = \alpha^2/K$ and behave as a  lattice gas of interacting hard-core classical charge $2e$ particles. (It is convenient to think of the bipolarons as hard-core bosons, but because they are non-dynamical in this limit, they in fact have no meaningful quantum statistics.)  The change in the nature of the low-energy states manifests as a pseudogap in the single-particle electron spectrum, onsetting at a temperature $T^\star \sim V$. At lower $T$  and for sufficiently large $\lambda_0$ we find a $\mathbf Q = (\pi,\pi)$ CDW state. While ME theory is extremely accurate for $T>T^\star$, we will see that it fails to describe the crossover at $T \sim T^\star$, and misses the strong-coupling physics when $T < T^\star$ entirely.  By contrast, the strong-coupling expansion (also discussed below) gives a satisfactory account of the system in the strong-coupling regime;  in particular, the CDW phase boundary labeled $T_c^\mathrm{Ising} \sim 1/\lambda_0$ in the figure was  computed to leading order in the strong-coupling expansion for the same parameters as in the DQMC study. Variations of the density with temperature and coupling strength are shown in Figure \ref{fig:density_vs_temp}.

\begin{figure*}[t!]
    \centering
   \includegraphics[width=\textwidth]{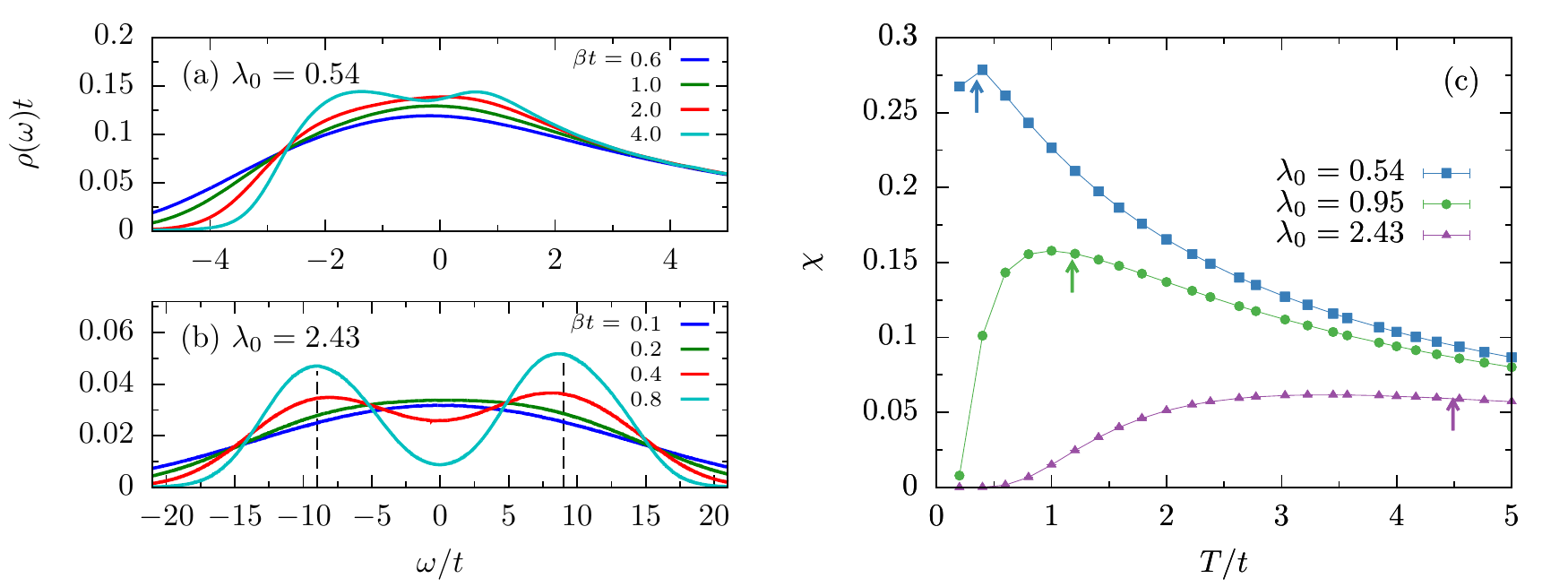}
    \caption{DOS as a function of energy relative to $E_F$ at various $T$ (a) 
    for $\lambda_0 = 0.54$ and (b) for $\lambda_0 = 2.43$. The dashed vertical lines indicate the bipolaron binding energy $\pm V$. Linear system size is $L=12$. Delta functions are broadened with $\eta = 10/L^2$.  $T^\star$ is identified as the highest temperature at which there is a local minimum at $\omega=0$.  (c) The charge or spin susceptibility as a function of $T$ for various $\lambda_0$;  the arrows indicated $T^\star$ inferred from the DOS. Linear system size is $L=10$.} 
    \label{fig:pseudogap}
\end{figure*}

We have previously carried out DQMC results for this model with large but finite $M$ such that $w=0.1$. We expect the thermodynamic properties for $w=0$ to be similar to those with $w=0.1$ when $T >\hbar \omega_0$ or to the right of the $T^\star(\lambda_0)$ line. While we have not explicitly tested this in all cases (especially at large $\lambda_0$ and low $T$, where the $w\neq 0$ DQMC is most difficult), we have verified the validity of this expectation wherever we have $w\neq 0$ results.  Even to the left of the $T^\star(\lambda_0)$ line, the results with $w=0$ and $w=0.1$ differ little down to temperatures that are a small fraction of $\hbar\omega_0$.  However, in this weak coupling regime, for finite $w$ (but still $w \ll 1$) we expect a SC transition at $T_c \sim \hbar\omega_0 \exp(-1/\lambda)$. This accounts for the one qualitative difference between Fig. \ref{schematic} and \ref{fig:phase_diag};  in the former we have added a SC phase below a critical temperature computed according to ME theory,\cite{PhysRevB.97.140501} while of course $T_c=0$ in the $M\to\infty$ limit.

\textit{CDW phase} -- At zero temperature, mean-field theory is exact in the $M \rightarrow \infty$ limit and gives a first order transition to a $\mathbf Q = (\pi, \pi)$ CDW state at a critical coupling strength $\lambda_0 = \bar\lambda_0\approx 0.31$, indicated by the orange circle in Fig. \ref{fig:phase_diag}. At fixed density, the system would phase separate into a region with $n=1$ and a region with $n<0.8$. For $\lambda_0> \bar \lambda_0$, the $(\pi, \pi)$ state persists to $T > 0 $ and  the finite temperature transitions we have observed appear continuous. (Presumably, this is not the case at low enough $T$ since,  on theoretical grounds, if there is a first order transition at $T=0$ one would expect it to persist to small non-zero $T$.) The CDW phase boundary is identified by finite-size scaling of the phonon correlation function at wave-vector $\mathbf Q$,
\begin{equation}
	D(\mathbf Q) = \frac{1}{L^2}\sum_{ij} e^{i \mathbf Q \cdot (\mathbf R_i - \mathbf R_j)}\langle x_i x_j\rangle,
\end{equation}
where $L$ is the linear system size, assuming the transition to be in the Ising university class. Details of this analysis are provided in the Appendix.

From a strong-coupling expansion in $1/\lambda_0$ one finds the effective Hamiltonian for the system (which gives a valid description for temperatures $T \ll V$) is an antiferromagnetic Ising model in a uniform external field: \cite{PhysRevB.48.6302, Carlson2008}
\begin{equation}
	H_\mathrm{eff} = \sum_{ij}J_{ij}\tau_i\tau_j - h \sum_i \tau_i,
	\label{eq:heff}
\end{equation}
where $\tau_i$ are classical Ising variables taking on the values $\pm 1$ and $J_{ij}=2t_{ij}^2/U$. The relation to the electronic degrees of freedom is that $\tau_i = 1$ if a site is occupied by a bipolaron and zero otherwise. The density $n$ of the original electrons and the magnetization $m$ of the Ising spins are related by $n=1+m$. In the parameter regime where the nearest-neighbor $J$ is much stronger than all further neighbor couplings and also near to half-filling ($m=0$), this model has a transition from a paramagnetic phase to a $(\pi,\pi)$ antiferromagnetic phase at a temperature $T_c^\mathrm{Ising} \sim J$. Depending on $m$ and the nature of the further neighbor couplings there may be additional ordering transitions or phase separation at lower temperatures.\cite{PhysRevB.21.1941} We have computed $T_c^\mathrm{Ising}$ for the parameters relevant to the model under consideration -- nearest-neighbor $J=2t^2/U$, next-nearest-neighbor $J'=2t'^2/U$, and external field $h$ tuned such that $m(T=0.25t)=-0.2$ -- and in Figure \ref{fig:phase_diag} we show that the transition temperature coincides very accurately with $T_\mathrm{cdw}$ of the full Holstein model for $\lambda_0 \gtrsim 1$. 

\textit{Pseudogap} -- The pseudogap region can be delimited by various crossover temperature scales. We define $T^\star(\lambda_0)$ as the temperature below which the electronic density of states (DOS) $\rho(\omega)$ develops a minimum at $\omega = 0$, similar to the conventional definition of the pseudogap temperature in various correlated materials. The DOS for a given phonon configuration $X$ is 
\begin{equation}
	\rho_X(\omega) = \frac{1}{L^2} \sum_\lambda \delta(\omega - E_\lambda [X]),
	\label{eq:dosx}
\end{equation}
where $E_\lambda$ are the single-particle energies in the phonon configuration $X$. The DOS is then obtained by averaging over phonon configurations (This procedure is explained in more detail in the Appendix). In practice the delta functions in \eqref{eq:dosx} are resolved with a Lorentzian broadening, with broadening parameter $\eta$ chosen to be on the order of the finite size gaps in the single-particle spectrum for a given system size. 
The DOS for representative weak and strong-coupling values is shown in Figure \ref{fig:pseudogap}.

The appearance of a pseudogap is also evident in thermodynamic observables. In the $w=0$ limit of the current model, the charge and spin susceptibilities are degenerate and we therefore define a single static susceptibility
\begin{equation}
\chi = \frac{\beta}{L^2}\left(\langle \hat N^2 \rangle - \langle \hat N \rangle^2\right) = \frac{\beta}{L^2}\left(\langle \hat M^2 \rangle - \langle \hat M \rangle^2\right), 
\end{equation}
were $\hat N = \sum_i \hat n_i$ and $\hat M = \sum_i \hat S_i^z$. In Figure \ref{fig:pseudogap} we see there is indeed a depression of $\chi$ below $T^\star$.

\begin{figure}[t!]
    \centering
    \includegraphics[]{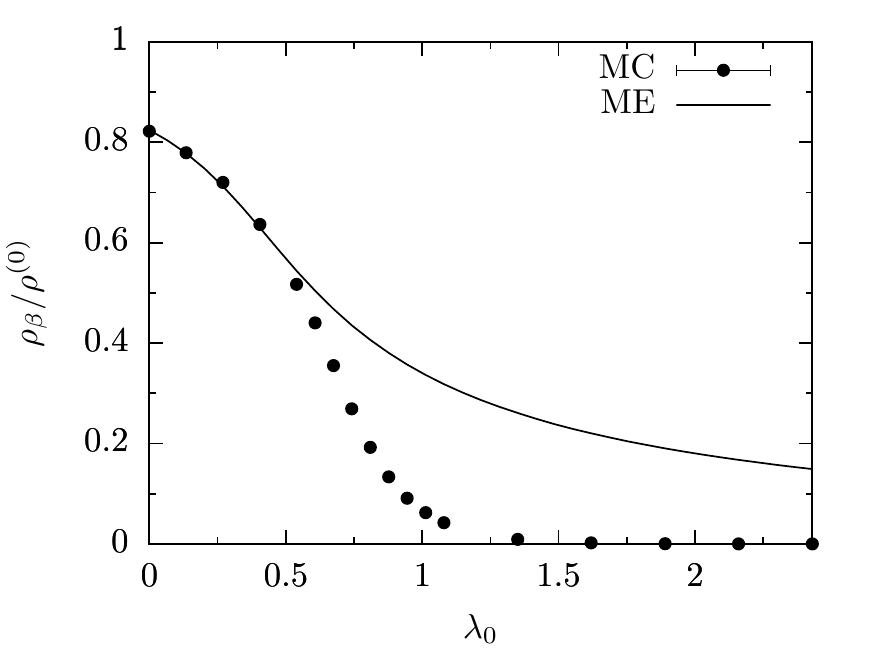}
    \caption{Temperature averaged DOS at the Fermi-energy $\rho_\beta$ at $T=0.25t$, normalized by the zero-temperature, non-interacting DOS $\rho^{(0)}$. At this elevated temperature $\rho_\beta < \rho^{(0)}$ even for $\lambda_0 = 0$. The breakdown of ME theory occurs for $\lambda_0 \approx 0.5$. Linear system size is $L=12$.}   
    \label{fig:gb2_beta4.0}
\end{figure}

\textit{ME theory} -- Migdal-Eliashberg (ME) theory purports to solve the electron-phonon problem for any coupling strength $\lambda_0$, provided the product $\lambda_0 w \ll 1$.\cite{migdal1958, eliashberg1960} For $w = 0$ the ME theory should therefore be valid for arbitrary $\lambda_0$. To assess the validity of this statement we compare the single-particle DOS computed within ME theory to that obtained with MC results. Our ME calculations are carried out in imaginary time and therefore comparison with dynamical quantities (e.g., single-particle DOS) requires analytic continuation. Rather than dealing with complications associated with analytic continuation we will work with a proxy for the low-energy DOS:
\begin{equation}
\rho_\beta \equiv \frac \beta\pi  G(\mathbf x=0, \tau = \beta/2) = \frac {\beta}{2\pi} \int d\omega \frac{\rho(\omega)}{\cos(\beta \omega/2)}.
\end{equation}
This quantity is essentially the single-particle DOS averaged over an energy window of order the temperature. At low-temperatures $\rho^{(0)} \approx \rho_\beta$. In Figure \ref{fig:gb2_beta4.0} we show $\rho_\beta$ computed with MC and within ME theory, at $T=0.25t$ where the density in both calculations is $n=0.8$. We find ME becomes qualitatively incorrect for $\lambda_0 \gtrsim 0.5$, where the MC shows a precipitous drop due to the onset of the pseudogap ($T^\star \approx 0.25t$ for $\lambda_0 = 0.5$), while the ME shows a much weaker dependence. We emphasize that this temperature is well above $T_\mathrm{cdw}$ ($T_\mathrm{cdw} \approx 0.1t$ for $\lambda_0 = 0.5$) and therefore the breakdown of ME theory is unrelated with the onset of CDW order. Indeed, we find that $T_\mathrm{cdw}$ drops rapidly as the density decreases from $n=1$, while $T^\star$ is essentially unchanged. Rather, the breakdown occurs because of a dramatic rearrangement in the low-energy spectrum upon entering the pseudogap regime. When $\lambda_0 \gtrsim 1$, and even when $\lambda_0 w = 0$, the ME perturbation theory breaks down because it is a perturbative expansion around the wrong state. 

\textit{Conclusion} -- While ME theory works well for sufficiently weak coupling, it breaks down to the right of the $T^\star(\lambda_0)$ line where the system is described by a classical lattice gas whose low-energy excitations are bipolarons. Because $T^\star$ exceeds significantly the ordering temperature $T_\mathrm{cdw}$ we do not associate the breakdown of ME theory with a competing order or fluctuations near $T_\mathrm{cdw}$. 


Interesting materials typically have multiple phonon branches, often multiple electronic bands crossing the Fermi energy, and generally more structured electron-phonon coupling, so quantitative comparison with the results for the Holstein model are of course not possible.  However, we feel that aspects of the present results are of general relevance.  Two aspects of the results, in particular, are relevant to phonon-mediated  superconductivity.  On the one hand,  the breakdown of ME theory  when $\lambda_0 \sim 1$ appears to be  unavoidable; for example, seeking ways to prevent a lattice instability (e.g. CDW ordering) does {\em not}, by itself, extend the range of validity of ME theory.  Moreover, while large $\lambda_0$ can indeed produce a large pairing scale, the resulting bipolaron formation is accompanied by a {\rm drop} in the superconducting susceptibility.  These results further corroborate our earlier inference\cite{esterlis2018bound} that there is an optimal value of $\lambda_0 \sim 1$ at which $T_c$ is maximal, and that $T_c$ always drops quickly to zero for larger $\lambda_0$.

In exploring whether the  optimal $\lambda_0$ obtained here is consistent with experimental data, it is important to distinguish  the bare value of the electron-phonon coupling -- our $\lambda_0$ -- from the renormalized value, $\lambda$, which can be extracted, for example, from tunneling data. The induced phonon softening that is prominent at larger coupling strength results in values of $\lambda>\lambda_0$.  For example, for the Holstein model with the same parameters studied here,\cite{PhysRevB.97.140501} $\lambda_0=0.5$ corresponds to  $\lambda \approx 2$.  In this context, we note that in the famous Allen-Dynes\cite{PhysRevB.12.905} compilation of experimental values of $\lambda$ and $T_c$ for a large number of conventional superconductors, all the entries are  roughly in the range $  \lambda\lesssim 2$.  The presence of an apparent upper bound on $\lambda$ in SCs is something that is not expected on the basis of ME theory.

 \noindent{\it Acknowledgements:} We would like to acknowledge helpful discussions with Yoni Schattner and Edwin Huang. SAK and IE were supported, in part, by NSF grant \# DMR-1608055 at Stanford. DJS was supported by the Scientific Discovery through Advanced Computing (SciDAC) program funded by U.S. Department of Energy, Office of Science, Advanced Scientific Computing Research and Basic Energy Sciences, Division of Materials Sciences and Engineering. Computational work was performed on the Sherlock computing cluster at Stanford University.
 %

%

\clearpage

\newpage

\appendix

\section{Monte-Carlo algorithm}

We study the Hamiltonian \eqref{eq:hamiltonian} in the $M\rightarrow \infty$ limit using a Monte Carlo (MC) technique. \cite{michielsen1996optical, PhysRevB.94.155150, PhysRevB.98.085405} In this limit it is useful to decompose the Hamiltonian as 
\be
H = H'_e  + H_p \label{eq:hamiltonian_Minf},
\ee
where
\begin{align}
H'_e &=   -\sum_{ij\sigma}t_{ij}c^\dag_{i\sigma}c_{j\sigma}  - \sum_{i\sigma}   (\mu - \alpha x_i)n_{i\sigma}, \\
H_p &= \frac K2 \sum_i x_i^2.
\end{align}
The partition function is
\be
Z = \mathrm{Tr} ~ e^{-\beta(H'_e + H_p)} = \int DX ~ e^{-\beta H_p[X]} Z_e[X] ,
\ee
where $\beta = 1/T$ is the inverse temperature, $X \equiv \left\{x_i \right\}$ denotes a configuration of the phonon fields, and $Z_e = \mathrm{Tr} \exp( -\beta H_e')$ is the electronic partition function in (classical) phonon configuration $X$. The trace for $Z_e$ can be computed explicitly and the result is
\begin{align}
	Z &= \int DX \det\left( 1 + e^{-\beta K[X]}\right)^2 e^{-\beta H_p[X]} \\
	&= \int DX ~ e^{-\beta H^\mathrm{eff}_p[X]},
	\label{eq:partition_function}
\end{align}
where $K$ is the bilinear form
\begin{equation}
	K_{ij} = -t_{ij} - (\mu - \alpha x_i)\delta_{ij},
\end{equation}
which is a functional of the phonon configuration, and
\be
H_p^\mathrm{eff} = H_p - 2\log \{\det[1 + \exp(-\beta K)] \}/\beta
\label{eq:hpeff}
\ee
is the effective Boltzmann weight. The partition function is classical -- i.e., the phonon configurations have no time dependence -- and is amenable to classical MC simulation.\cite{michielsen1996optical, PhysRevB.94.155150, PhysRevB.98.085405} Matrix $K$ is Hermitian, implying that its eigenvalues are real and hence the determinant appearing in \eqref{eq:hpeff} is non-negative. Therefore, there is no sign problem. Let $N$ denote the number of lattice sites. To perform the MC, the  $N\times N$ matrix $K$ must be diagonalized for each configuration $X$. The diagonalization step is the mostly costly and makes the computational time scale as $N^3$. 

Thermal expectation values $\langle \mathcal O \rangle$ are computed according to 
\be
\langle \mathcal O \rangle = \frac 1Z \mathrm{Tr} ~ \mathcal O e^{-\beta(H'_e + H_p)} = \frac 1Z  \int DX ~ e^{-\beta H_p^\mathrm{eff}[X]} \mathcal O[X],
\ee
where 
\be
\mathcal O[X] = \frac{1}{Z_e}\mathrm{Tr} ~ \mathcal O e^{-\beta H'_e}.
\ee
The functional $\mathcal O[X]$ is easily computed given the single-particle spectrum of $H_e'$.

To accelerate convergence to the thermodynamic limit, the calculations presented in this paper have been carried out in the presence of weak, uniform magnetic field.\cite{PhysRevB.65.115104}

\section{Zero temperature mean-field theory} 

\begin{figure}[t]
    \centering
    \includegraphics[width=0.5\textwidth]{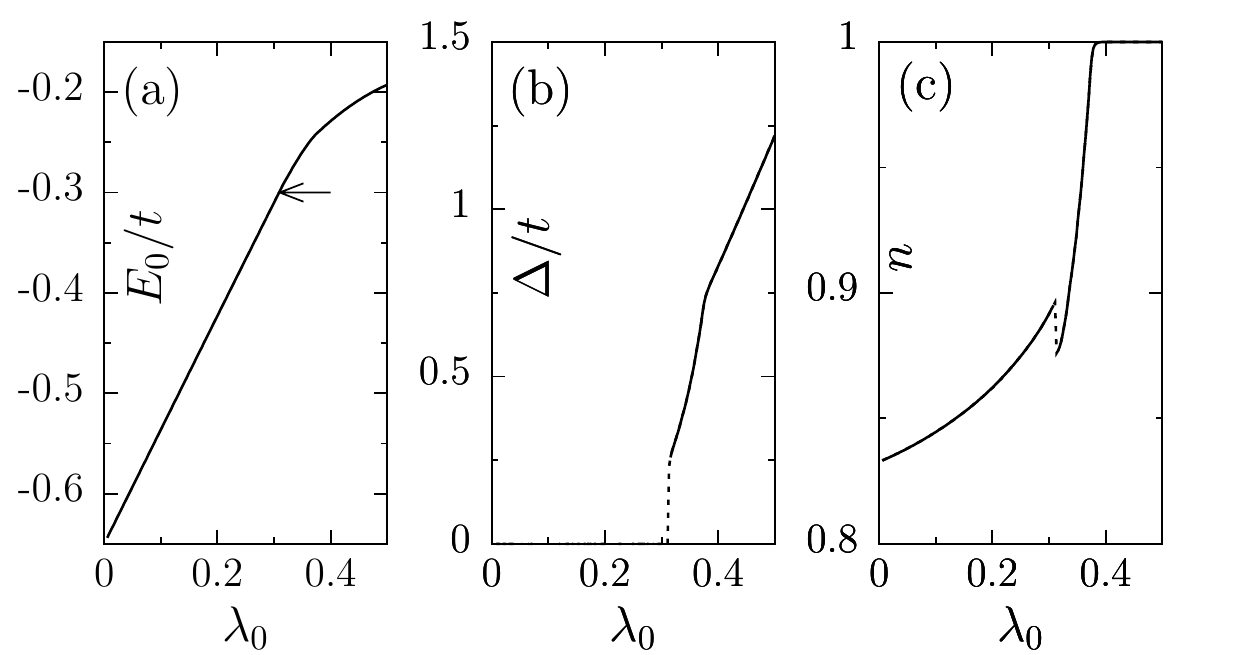}
    \caption{(a) Ground-state energy $E_0$ as a function of $\lambda_0$. The arrow indicates the first-order transition where $\Delta$ onsets. (b) Order parameter $\Delta$ as a function of $\lambda_0$; onsets discontinuously at $\lambda_0 = \bar\lambda_0 \approx 0.31$.  (c) Density $n$ as a function of $\lambda_0$. There is a discontinuity at $\bar \lambda_0$ and for $\lambda_0 > \bar\lambda_0$, $n$ rises continuously to one.}
    \label{fig:mft_figs}
\end{figure}

At $T=0$, where thermal fluctuations of the phonons are absent, mean-field theory is exact in the $M \rightarrow \infty$ limit. The ground-state energy of the Hamiltonian \eqref{eq:hamiltonian_Minf} is
\be
E_0[X] = \varepsilon_0[X] + \frac K2 \sum_i x_i^2,
\ee
where $\varepsilon_0$ is the ground-state energy of $H'_e$ in phonon configuration $X$. The ground-state phonon configuration is obtained by minimizing $E_0$:
\be
\frac{\partial E_0}{\partial x_i} = \frac{\partial \varepsilon_0}{\partial x_i} + K x_i = 0.
\ee 
By the Feynman-Hellmann theorem, $\partial \varepsilon_0 / \partial x_i = \alpha \langle n_i \rangle_0$, where $\langle \ldots \rangle_0$ is the ground-state expectation value. This yields the self-consistency condition
\be
x_i = -\frac{\alpha}{K} \langle n_i \rangle_0.
\label{eq:self_consis_x}
\ee
Solving the $N$ equations in \eqref{eq:self_consis_x} for $X$ gives the ground-state phonon configuration. 

From our finite $T$ DQMC studies, we see that the ground-state for the range of parameters studied here is always either translationally invariant, or has $(\pi,\pi)$ CDW order.  The $T=0$ calculation is greatly simplified if we take this as justification to assume that in the ground-state $x_i = \bar x + (-1)^{i_x + i_y} \delta x$. In this case the Hamiltonian \eqref{eq:hamiltonian_Minf} takes the following form in momentum-space:
\be
\begin{aligned}
H &= \sum_\mathbf k [\epsilon_\mathbf k - (\mu + \bar \mu)]c^\dag_{\mathbf k \sigma}c_{\mathbf k\sigma}  \\
&\quad + \Delta \sum_\mathbf k c^\dag_{\mathbf k + \mathbf Q,\sigma} c_{\mathbf k + \mathbf Q,\sigma} \\
&\quad +  \frac{1}{2V}\bar \mu^2 N + \frac{1}{2V}\Delta^2 N.
\end{aligned}
\ee
where $\bar \mu = \alpha \bar x$, $\Delta = \alpha \delta x$, and $V = \alpha^2/K$. Rather than the full set of phonon coordinates there are now only two variational parameters $\bar \mu$ and $\Delta$. The self-consistency condition \eqref{eq:self_consis_x} can now be written as two coupled equations:
\be
\bar \mu = -V\langle n \rangle_0, \quad \Delta = -V \langle n_\mathbf Q \rangle_0.
\label{eq:self_consis_simp}
\ee
To compute the $T=0$ portion of the phase diagram in Fig. \ref{fig:phase_diag} we solve Equations \eqref{eq:self_consis_simp} with $t'/t=-0.3$ and $\mu = \mu(\lambda_0)$ such that $n(T=0.25t) = 0.8$. The function $\mu(\lambda_0)$ is obtained from the finite $T$ MC calculations. In Figure \ref{fig:mft_figs} we show the ground-state energy, $E_0$, order parameter $\Delta$, and density $n$ as a function of $\lambda_0$. The transition to the $(\pi,\pi)$ state is first-order, as can be seen from the discontinuous onset of $\Delta$ for $\lambda_0 = \bar \lambda_0 \approx 0.31$.
A second critical point can be seen at $\lambda_0\approx 0.39$, beyond which $n = 1$.

\section{Determination of $T_\mathrm{CDW}$}

\begin{figure}[t!]
    \centering
    \includegraphics[width=0.5\textwidth]{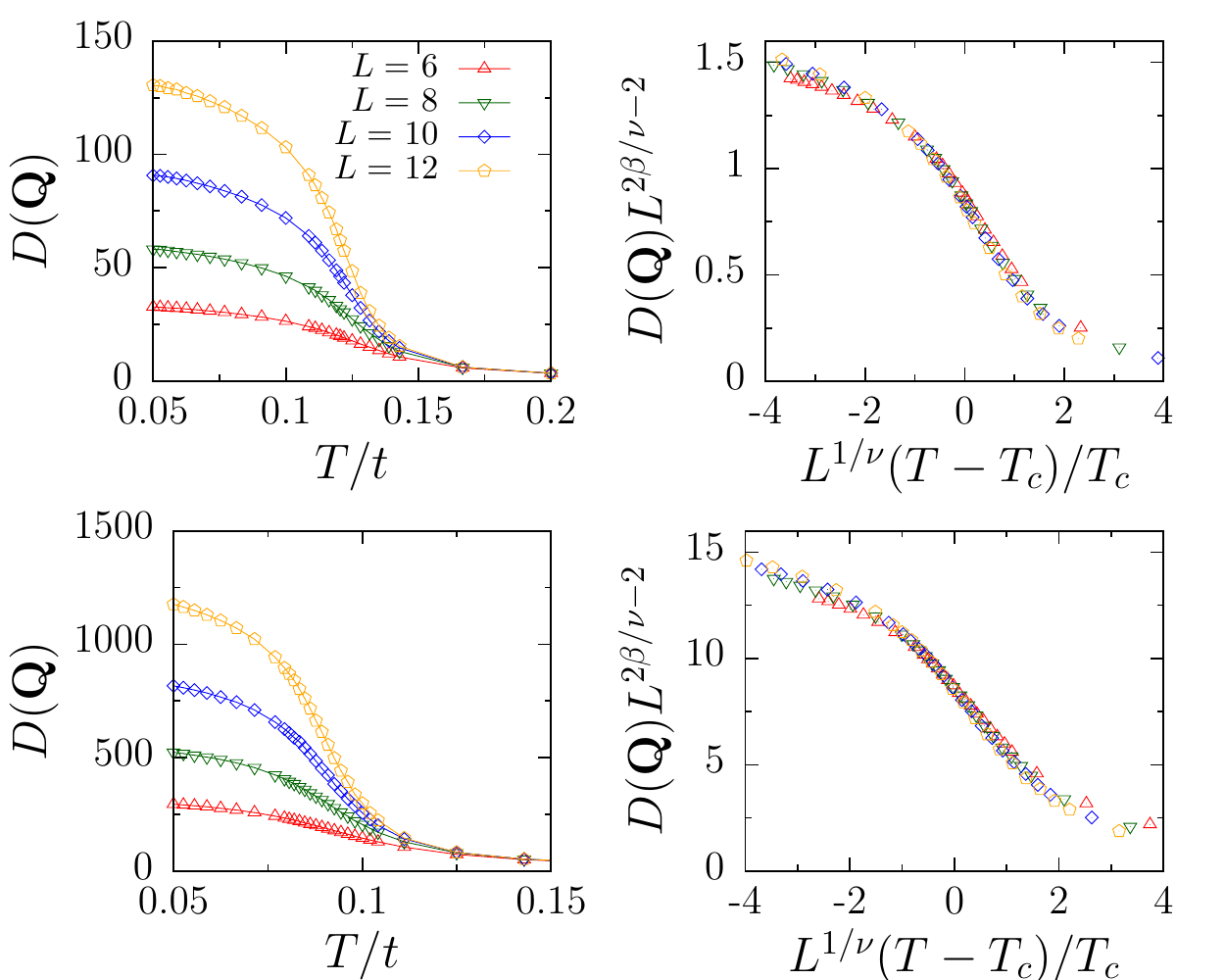}
    \caption{Finite-size scaling of phonon correlation function $D(\mathbf Q)$, with $\mathbf Q = (\pi,\pi)$ for (a) $\lambda_0 = 0.54$ and $T_\mathrm{cdw} \approx 0.12t$ and (b)  $\lambda = 2.43$ and $T_\mathrm{cdw} \approx 0.09t$. Ising exponents $\nu = 1$ and $\beta = 1/8$ are used.}
    \label{fig:fss}
\end{figure}

\begin{figure}[t!]
    \centering
    \includegraphics[width=0.5\textwidth]{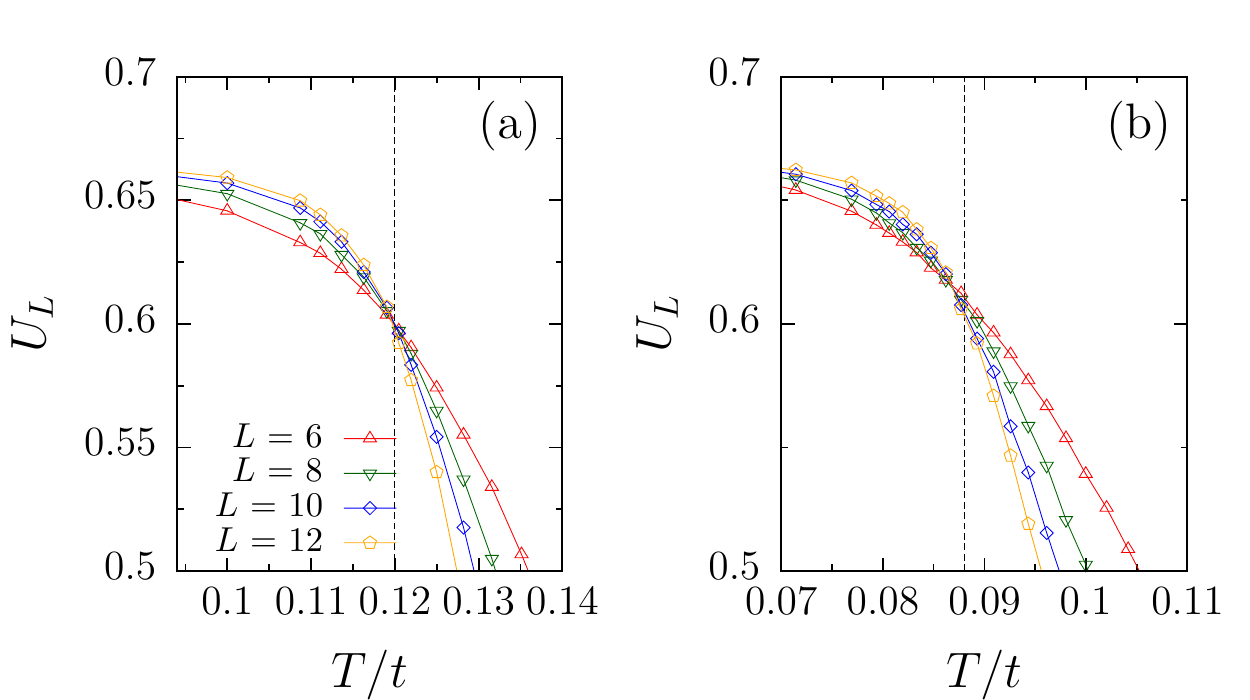}
    \caption{System size dependence of the Binder cumulant $U_L$ for (a) $\lambda_0 = 0.54$ and (b) $\lambda_0 = 2.43$. Vertical dashed lines indicate $T_\mathrm{cdw}$ obtained from finite-size scaling of $D(\mathbf Q)$}
    \label{fig:opbc}
\end{figure}

\begin{figure}[t]
    \centering
    \includegraphics[width=0.5\textwidth]{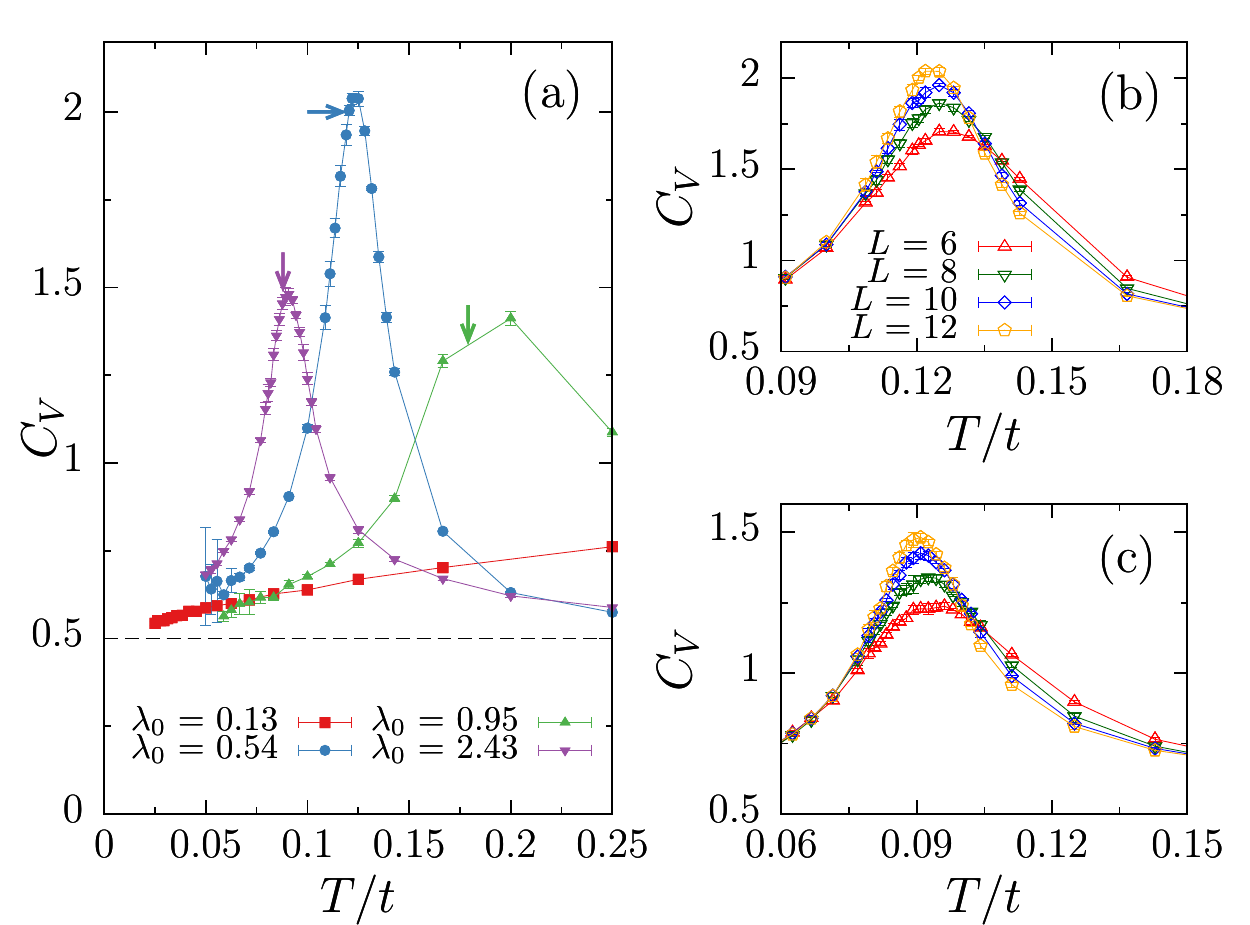}
    \caption{(a) Specific heat $C_V$ as a function of temperature $T$ for various coupling strengths. The dashed line indicates the contribution from non-interacting phonons. Arrows show the $T_\mathrm{cdw}$ inferred from finite-size scaling. Linear system size is $L=12$. Finite size scaling of $C_V$ for (b) $\lambda_0 = 0.54$ and (c) $\lambda_0 = 2.43$.}
    \label{fig:c_fss}
\end{figure}

The CDW transition temperature $T_\mathrm{cdw}$ is obtained from the finite size scaling behavior of the phonon correlation function
\begin{equation}
	D(\mathbf q) = \frac{1}{L^2}\sum_{ij} e^{i \mathbf q \cdot (\mathbf R_i - \mathbf R_j)}\langle x_i x_j\rangle
\end{equation}
at wave-vector $\mathbf q = \mathbf Q = (\pi,\pi)$. Near the transition $D(\mathbf Q)$ has the scaling form
\begin{equation}
	\frac{1}{L^2}D(\mathbf Q) = L^{-2\beta/\nu}f[L^{1/\nu}(T - T_c)/T_c].
\end{equation}
The $(\pi,\pi)$ transition spontaneously breaks $Z_2$ symmetry and we therefore use exponents for the 2D Ising universality class: $\nu = 1$ and $\beta = 1/8$. The transition temperature $T_\mathrm{cdw}$ is obtained by looking for the best data collapse. This procedure is shown in Figure \ref{fig:fss} for representative weak and strong-coupling values, $\lambda_0 = 0.54$ and $\lambda_0 = 2.43$. 

The value of $T_\mathrm{cdw}$ obtained from finite-size scaling is consistent with the expected behavior of the Binder cumulant of the order parameter and the specific heat near a continuous phase transition. We take the order parameter to be the staggered phonon displacement
\begin{equation}
\phi = \frac{1}{L^2}\sum_i (-1)^{i_x + i_y} x_i 
\end{equation}
and define the Binder cumulant $U_L$\cite{binder1981finite} in the usual way
\begin{equation}
U_L = 1 - \frac{\langle \phi^4\rangle}{3\langle \phi^2 \rangle^2}.
\end{equation}
The Binder cumulant has the property that, in the thermodynamic limit, $U_L \rightarrow 0$ for $T > T_\mathrm{cdw}$ and $U_L \rightarrow 2/3$ for $T < T_\mathrm{cdw}$. The intersection point of $U_L$ for different system sizes gives an estimate of $T_\mathrm{cdw}$. In Figure \ref{fig:opbc} we show $U_L$ in the weak and strong-coupling regimes. The values of $U_L$ for different system sizes indeed intersect at temperatures near $T_\mathrm{cdw}$ obtained from finite-size scaling.

In Figure \ref{fig:c_fss} we show the specific heat $C_V$ as a function of $T$ for several values of $\lambda_0$. The $(\pi,\pi)$ CDW transition is in the 2D Ising universality class, for which the specific heat exponent is known to be $\alpha = 0$, implying a logarithmic divergence of the specific heat at $T_\mathrm{cdw}$ upon approaching the thermodynamic limit $C_V \sim C_V^0 \ln L$. While we do not access sufficiently large systems to see scaling behavior in the specific heat, Fig. \ref{fig:c_fss} shows the position of the specific heat maxima is consistent with our estimates of $T_\mathrm{cdw}$. Furthermore, the evolution of the peak positions with system size tends toward the estimated $T_\mathrm{cdw}$, as can be seen in Fig. \ref{fig:c_fss}.

\end{document}